\documentclass[aps,prb,twocolumn]{revtex4}
\usepackage{amssymb}
\usepackage{amsmath}
\usepackage{graphicx}
\usepackage{epsfig}
\usepackage{subfigure}

\begin{document}

\title{One-step multi-qubit GHZ state generation in a circuit QED system}
\author{Ying-Dan Wang, Stefano Chesi, Daniel Loss, Christoph Bruder}
\affiliation{Department of Physics, University of Basel,
Klingelbergstrasse 82, 4056 Basel, Switzerland}

\begin{abstract}
We propose a one-step scheme to generate GHZ states for
superconducting flux qubits or charge qubits in a circuit QED setup.
The GHZ state can be produced within the coherence time of the
multi-qubit system. Our scheme is independent of the initial state
of the transmission line resonator and works in the presence of
higher harmonic modes. Our analysis also shows that the scheme is
robust to various operation errors and environmental noise.
\end{abstract}


% 03.67.Bg Entanglement production and manipulation
% 85.25.Cp Josephson devices
% 03.67.Lx Quantum computation architectures and implementations

\maketitle

Entanglement is the most important resource for quantum information
processing. Therefore, the question of how to prepare maximally
entangled states, i.e., the GHZ state, or the Bell states in the
two-qubit case, in various systems remains an important issue.
Superconducting Josephson junction qubits are one of the promising
solid-state candidates for a physical realization of the building
blocks of a quantum information processor, see
e.g.~\cite{Makhlin2001,YouT2005,Wendin2006,Clarke2008}. They are
undergoing rapid development experimentally, in particular, in
circuit QED setups. Two-qubit Bell states have been demonstrated
experimentally~\cite{Steffen2006,Plantenberg2007,Filipp2008}.  There
are also some theoretical proposals on how to generate maximally
entangled states for two or three
qubits~\cite{Plastina2001,Wei2006,Bodoky2007,Kim2008,Zhang2009,Galiautdinov2008,Beat2008,Beat2009,Hutchison2009}.
However, how to scale up to multi-qubit GHZ state generation remains
an open question. Some general schemes based on fully connected
qubit network is proposed but no specific circuit design is
provided~\cite{Galiautdinov2009}. Most recently, preparation of
multi-qubit GHZ states was proposed based on
measurement~\cite{Helmer2009,Bishop2009}. This type of state
preparation is probabilistic and the probability to achieve a GHZ
state decreases exponentially with the number of qubits. In this
paper, we propose a GHZ state preparation scheme based on the
non-perturbative dynamic evolution of the qubit-resonator system.
The preparation time is short and the preparation is robust to
environmental decoherence and operation errors.

\section{The coupled circuit QED system}

\begin{figure}[tp]
\begin{center}
\includegraphics[bb=134 335 461 628, scale=0.6, clip]{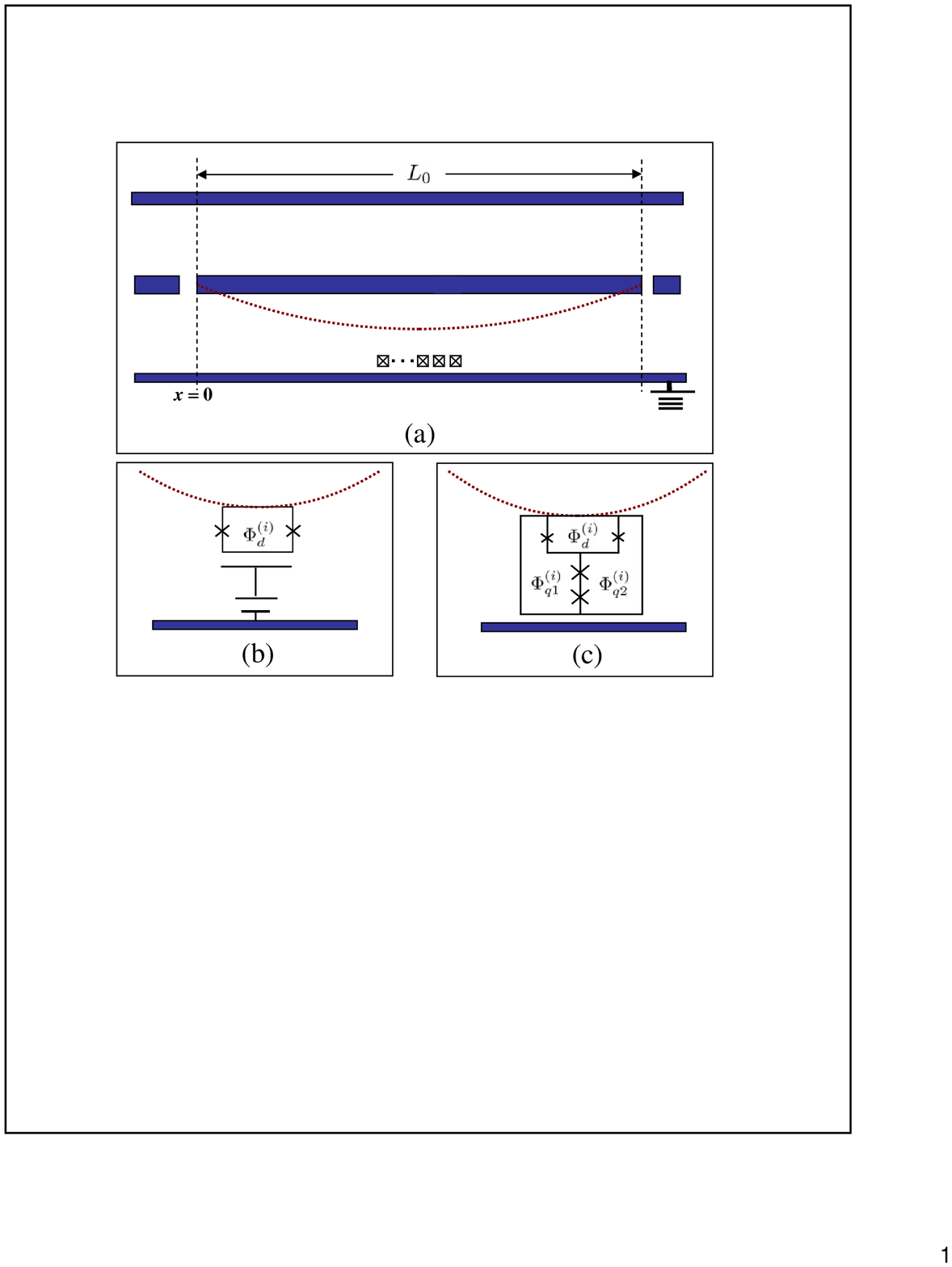}
\end{center}
\caption{(Color online) Schematic diagram of our setup. (a) The
qubits are coupled through a superconducting stripline resonator
(the blue stripe). Each `crossed box' denotes one qubit which can be
either a charge qubit or a flux qubit; the dashed red line shows the
magnitude of the magnetic field. (b) Detailed schematic of a charge
qubit. The crosses denote Josephson junctions; (c) Detailed
schematic of a gradiometer-type flux qubit. The crosses denote
Josephson junctions. } \label{fig:circuit}
\end{figure}
The GHZ state preparation scheme described below is based on a circuit
QED setup where superconducting qubits are strongly coupled to a 1D
superconducting transmission line resonator (TLR).
Figure~\ref{fig:circuit}(a) shows the type of circuit we have in mind:
a qubit array is placed in parallel with a line of length $L_{0}$. The
superconducting transmission line is essentially an LC resonator with
distributed inductance and
capacitance~\cite{Blais2004,Wallraff2004}. The oscillating
supercurrent vanishes at the end of the transmission line and this
provides the boundary condition for the electromagnetic field of this
on-chip resonator. The qubits are fabricated around the central
positions $x= L_0/2$. Since the qubit dimension (several micrometer)
is much smaller than the wave length of the fundamental
electromagnetic modes (centimeter), the coupling between the qubits
and the TLR is approximately homogeneous. Since $x=L_0/2$ is an
antinode of the magnetic field where the electric field is zero,
the qubits are only coupled to the magnetic component, which induces
a magnetic flux $\Phi'$ through the superconducting loop given by
\begin{equation}
\Phi'=\eta^{(i)} \frac{\Phi_{0}}{\pi}(a+a^{\dag})
\end{equation}
with
\begin{equation}
\eta^{(i)}
=\frac{M^{(i)}\pi}{\Phi_{0}}\sqrt{\frac{\hbar\omega}{2L}}\; .
\label{eta}
\end{equation}
Here, $M^{(i)}$ is the mutual inductance between the resonator and
the $i$-th qubit, $\omega=\pi/(LC)^{1/2}$ is the frequency of the
fundamental resonator mode, $\Phi_{0}=h/2e$ is the magnetic flux
quantum and $L$ ($C$) is the total self-inductance (capacitance) of
the stripline. Here, we have assumed the qubit array to be only
coupled with a single mode of the resonator, and $a$ ($a^{\dag}$) is
the annihilation (creation) operator of this fundamental mode.

The stripline resonator can be used to couple both charge qubits and
flux qubits as described below.

\subsection{Charge qubit system}

We first consider the charge qubit case. Suppose each qubit is a
charge qubit (see Fig.~\ref{fig:circuit}(b)) consisting of a
dc-SQUID formed by a superconducting island connected to two
Josephson junctions. The Coulomb energy of each qubit is modified by
an external bias voltage and the effective Josephson tunneling
energy is determined by the magnetic flux $\Phi_{x}^{(i)}$ threading
the dc-SQUID. The Hamiltonian of a single charge qubit
reads~\cite{Nakamura1999}
\begin{equation}
H^{(i)}\equiv \frac{E_{C}^{(i)}}{4}(1-2n_{g}^{(i)}) \sigma_{z}^{(i)}
-E_{J}^{(i)}\cos(\pi\frac{\Phi_{x}^{(i)}}{\Phi_{0}})
\sigma_{x}^{(i)}\; , \label{h0}
\end{equation}
where $E_{c}^{(i)}$ ($E_{J}^{(i)}$) is the Coulomb
(Josephson) energy of the $i$-th qubit, and $n_{g}^{(i)}$ is the bias
charge number that can be controlled by an external gate voltage. The Pauli
matrices
$\sigma_{z}=|0\rangle \left\langle 0\right\vert
-|1\rangle \left\langle 1\right\vert \,$, $\sigma_{x}=|0\rangle
\left\langle 1\right\vert |1\rangle \left\langle 0\right\vert \,$ are
defined in terms of the charge eigenstates
$\left\vert 0\right\rangle$ and
$\left\vert 1\right\rangle $. $\left\vert 0\right\rangle $ and $\left\vert
1\right\rangle $ denote $0$ and $1$ excess Cooper pair on the island
respectively. $\Phi_{d}^{(i)}=\Phi_{e}^{(i)}+\Phi'^{(i)}$
includes contributions from both the
external flux bias $\Phi_{e}^{(i)}$ and the flux $\Phi'^{(i)}$.

For small $\eta^{(i)}$, the Josephson energy can be expanded to
linear order in $\eta^{(i)}$, which results in an additional linear
coupling between the $x$-component of the qubits and the bosonic
mode. If all the qubits are assumed to be biased at the degeneracy
point $n_{g}^{(i)}=1/2$, the total Hamiltonian reads
\begin{equation}
H=\sum_{i}\left(\Omega^{(i)}(\Phi_{e}^{(i)})\sigma_{x}^{(i)}+
g^{(i)}(\Phi_{e}^{(i)}) (a+a^{\dag}) \sigma_{x}^{(i)}\right)
+H_{LC}\; , \label{ch}
\end{equation}
with the single charge qubit energy splitting
$\Omega^{(i)}(\Phi_{e}^{(i)}) =
-E_{J}^{(i)}\cos(\pi\Phi_{e}^{(i)}/\Phi_{0})$,
$g^{(i)}(\Phi_{e}^{(i)})=\eta^{(i)}E_{J}^{(i)}\sin(\pi\Phi_{e}^{(i)}/\Phi_{0})$,
and the free Hamiltonian of the TLR $H_{LC}=\omega a^{\dag}a$. Note
that the coupling between the qubits and the TLR can be turned off
by setting $\Phi_{e}^{(i)}=n\Phi_{0}$.

\subsection{Flux qubit system}

For a flux qubit system, a circuit example to realize our proposal
is shown in Fig.~\ref{fig:circuit}(c). The $i$-th qubit contains
four Josephson junctions in three loops instead of one or two loops
in the conventional flux qubit design~\cite{Mooij1999,Orlando1999}.
The two junctions in the dc-SQUID have identical Josephson energies
$\alpha_{0}^{(i)}E_{J}^{(i)}$, here $\alpha_{0}^{(i)}$ is the ratio
between the Josephson energy of the smaller junction and that of
the two bigger junctions~\cite{Mooij1999,Orlando1999}. The other two
junctions are assumed to have the Josephson energy $E_{J}^{(i)}$.
The superconducting loops are penetrated by magnetic fluxes
$\Phi_{q1}^{(i)}$, $\Phi_{q2}^{(i)}$, and $\Phi_{d}^{(i)}$
respectively. The corresponding phase relations are
\begin{eqnarray}
\varphi_{4}^{(i)}-\varphi_{3}^{(i)} &=&2\pi
\Phi_{d}^{(i)}/\Phi_{0} \\
\varphi_{1}^{(i)}+\varphi_{2}^{(i)}
+\frac{\varphi_{3}^{(i)}}{2}+\frac{\varphi_{4}^{(i)}}{2}
&=&2\pi (\Phi_{q1}^{(i)}-\Phi_{q2}^{(i)})/\Phi_{0} \\
\Phi_{q1}^{(i)}+\Phi_{q2}^{(i)}+\Phi_{d}^{(i)} &=&n\Phi_{0}\; ,
\end{eqnarray}
where $\varphi_{k}$ ($k=1,2,3,4$) is the phase difference across the $k$-th
junction. The total Josephson energy of the circuit is
\begin{align}
-U_{0}^{(i)}&=E_{J}^{(i)}\cos \varphi_{1}^{(i)}+ E_{J}^{(i)}\cos
\varphi_{2}^{(i)}\nonumber\\
&+\alpha^{(i)}E_{J}^{(i)}\cos\left(2\pi\Phi_{t}^{(i)}/\Phi_{0}-
(\varphi_{1}^{(i)}+\varphi_{2}^{(i)})\right)
\end{align}
with $\Phi_{t}^{(i)}\equiv \Phi_{q1}^{(i)}-\Phi_{q2}^{(i)}$ and
$\alpha^{(i)}=2\alpha_{0}^{(i)}\cos(\pi\Phi_{d}^{(i)}/\Phi_{0})$.
If $\Phi_{t}^{(i)}$ is biased close to $\Phi_{0}/2$, the circuit
becomes a flux qubit, i.e., a two-level system in the quantum
regime~\cite{Mooij1999,Orlando1999}. Together with
the charging energy, the total Hamiltonian for the $i$-th qubit is
\begin{equation}
H^{(i)}=\varepsilon^{(i)}(\Phi_{t}^{(i)})
\sigma_{z}^{(i)}+\Delta^{(i)}(\Phi_{d}^{(i)}) \sigma_{x}^{(i)}\; .
\end{equation}
The Pauli matrices read $\sigma_{z}=|0\rangle\langle 0\vert -
|1\rangle\langle 1\vert$, $\sigma_{x}=|0\rangle\langle 1\vert
|1\rangle \langle 0\vert$, and are defined in terms of the classical
current where $\vert 0\rangle$ and $\vert 1\rangle $ denote the
states with clockwise and counterclockwise currents in the loop. The
energy spacing of the two current states is
$\varepsilon^{(i)}(\Phi_{t}^{(i)}) \equiv
I_{p}^{(i)}(\Phi_{t}^{(i)}-\Phi_{0}/2)$, and the tunneling matrix element
between the two states is
$\Delta^{(i)}(\Phi_{d}^{(i)})\equiv \Delta^{(i)}(\alpha^{(i)})$.
Note that in contrast to the original flux qubit
design~\cite{Mooij1999,Orlando1999}, this gradiometer flux qubit is
insensitive to homogeneous fluctuations of the magnetic
flux~\cite{Paauw2009}. More importantly, it enables the TLR
to couple with the dc-SQUID loop without changing the total bias
flux of the qubit. As in the case of the charge qubit, the magnetic
flux in the dc-SQUID loop includes two parts:
$\Phi_{d}^{(i)}=\Phi_{e}^{(i)}+\Phi'^{(i)}$, where $\Phi_{e}^{(i)}$
is due to the external control line and $\Phi'^{(i)}$ is due to the
TLR.

For $\eta^{(i)} \ll 1$, one can expand the Hamiltonian in terms of
$\eta$.  The second-order terms $\sim \eta^{(i)2}d^2\Delta/d\alpha^2$
are much smaller than the zeroth and the first-order term.
The Hamiltonian of each qubit can be written as~\cite{Wang2009}
\begin{equation}
H^{(i)}=\varepsilon^{(i)}(\Phi_{t}^{(i)})
\sigma_{z}^{(i)}+\Delta^{(i)} (\Phi_{e}^{(i)})
\sigma_{x}^{(i)}+g^{(i)}(\Phi_{e}^{(i)}) \sigma_{x}^{(i)}(a+a^{\dag})\; .
\end{equation}
The coupling coefficient is
\begin{equation}
g^{(i)}(\Phi_{e}^{(i)}) =
-2\alpha_{0}^{(i)}\eta^{(i)}\sin(\pi\Phi_{e}^{(i)}/\Phi_{0})
\left.\frac{d\Delta(\alpha^{(i)})}{d\alpha^{(i)}}
\right\vert_{\Phi_{d}^{(i)}=\Phi_{e}^{(i)}}\; .
\end{equation}
Therefore, by setting $\Phi_{e}^{(i)}=n\pi$, the qubit-resonator
interaction can be turned off. When the interaction is on,
$\Phi_{e}^{(i)}$ can be tuned to compensate the difference of the
fabrication parameters and realize a homogeneous coupling
$g^{(i)}=g$. Then if each qubit is biased at the degeneracy point
$\Phi_{t}^{(i)}=(n+1/2)\Phi_{0}$, the total Hamiltonian becomes
\begin{equation}
H=\sum_{i}\Omega^{(i)}(\Phi_{e}^{(i)})\sigma
_{x}^{(i)}+g^{(i)}(\Phi_{e}^{(i)})\sigma_{x}^{(i)}(a+a^{\dag})
+H_{LC}\; , \label{ha}
\end{equation}
where $\Omega^{(i)}(\Phi_{e}^{(i)})=\Delta^{(i)}(\Phi_{e}^{(i)})$ is the
single qubit energy splitting. Comparing Eqs.~(\ref{ch}) and
(\ref{ha}), it is evident that the two Hamiltonians have the same
structure: the interaction term commutes with the free term, and the
interaction can be switched on and off. In the next section, we show
how to generate a multi-qubit GHZ state by utilizing these features.

\section{Generation of a GHZ state}

In the interaction picture,
\begin{equation}
H_{I}(t) =\sum_{i}g^{(i)}(a^{\dag}e^{i\omega t}+ ae^{-i\omega
t})\sigma_{x}^{(i)}\; . \label{hi}
\end{equation}
Since
$\{\sigma_{x}^{(i)}\sigma_{x}^{(j)},a\sigma_{x}^{(i)},a^{\dag}\sigma_{x}^{(i)},1\}$
form a closed Lie Algebra, the time evolution operator in the
interaction picture can be written in a factorized way
as~\cite{Wei1963}
\begin{align}
U_{I}(t)=&\prod_{i\neq
j}e^{-iA_{ij}(t)\sigma_{x}^{(i)}\sigma_{x}^{(j)}}
\prod\limits_{i}e^{-iB_{i}(t)a\sigma_{x}^{(i)}}\nonumber\\
&\times\prod\limits_{i}e^{-iB^{*}_{i}(t)a^{\dag}\sigma_{x}^{(i)}}e^{-iD(t)}\;
, \label{u1}
\end{align}
and $U_{I}(t)$ satisfies
\begin{equation}
i(\frac{\partial}{\partial t}U_{I}(t))U_{I}^{-1}(t)=H_{I}(t)\; .
\end{equation}
Solving this equation for the initial condition
$A_{ij}(0)=B_{i}(0)=D(0)=0$,
we obtain
\begin{align}
B_{i}(t)&=\frac{ig^{(i)}}{\omega}( e^{-i\omega t}-1) \\
A_{ij}(t)&=\frac{g^{(i)}g^{(j)}}{\omega}\left(\frac{1}{i\omega}
(e^{i\omega t}-1)-t\right) \label{ca}\\
D(t)&=\sum_{i}\frac{(g^{(i)})^{2}}{\omega}\left( \frac{1}{i\omega}
(e^{i\omega t}-1) -t\right)\; . \label{co}
\end{align}
In the Schr\"odinger picture
\begin{equation}
U_{s}(t)=U_{0}(t) U_{I}(t) =e^{-i\omega a^{\dag}at}
\prod_{i}e^{-i\Omega^{(i)}\sigma_{x}^{(i)}t}U_{I}(t)\; .
\end{equation}

Note that $B_{i}(t)$ is a periodic function of time and vanishes at
$t=T_{n}={2\pi n}/\omega$ for integer $n$. At these
instants of time, the time evolution operator takes the form
\begin{equation}
U(T_n) =\exp(-i\sum_{i\neq j}\theta_{ij}(n)
\sigma_{x}^{(i)}\sigma_{x}^{(j)}) \exp(-iD(t))\; , \label{ut}
\end{equation}
in the interaction picture. Here, $\theta_{ij}(n)=
g^{(i)}g^{(j)}T_n/\omega=g^{(i)}g^{(j)}2\pi n/\omega^2$. Thus, at
these times, the time evolution is equivalent to that of a system of
coupled qubits with an interaction Hamiltonian of the form $\propto
\sigma_{x}^{(i)}\sigma_{x}^{(j)}$. Therefore, by choosing
appropriate coupling pulse sequences, an effective XX-coupling can
be realized for multiple qubits. This coupling can be utilized to
construct a CNOT gate for two qubits~\cite{Wang2009}. If the
couplings are homogeneous for all qubits, i.e., $g^{(i)}=g$ (for
$i=1,..,N$),
\begin{equation}
\theta_{ij}(n)\equiv\theta(n)=\frac{g^2}{\omega^2}2\pi n\; ,
\label{theta}
\end{equation}
Eq.~(\ref{ut}) can be written as
\begin{equation}
U(T_n) =\exp(-i4\theta(n) J_{x}^{2}) \exp(i\theta(n) N)\exp(-iD(t))
\label{u2}
\end{equation}
with $J_{x}=\sum_{i}\sigma_{x}^{(i)}/2$.

Suppose the initial state of the qubits is
\begin{equation}
\left\vert\Psi(0) \right\rangle
=\bigotimes_{i=1}^{N}\left\vert -\right\rangle_{z}^{(i)}
\end{equation}
where $\vert\pm\rangle_{z}$ denotes the eigenstates of $\sigma_{z}$,
$\sigma_{z}\vert\pm\rangle_{z}=\pm\vert\pm\rangle_{z}$. This initial
state can be prepared by biasing the qubits far away from the
degeneracy point, letting them relax to the ground state and then
biasing them back adiabatically. Starting from the initial state,
under the time evolution described by Eq.~(\ref{u2}), the state
evolves into a GHZ state~\cite{Molmer1999,You2003b} (up to a global
phase factor)
\begin{equation}
\left\vert \Psi (T_n) \right\rangle =\frac{1}{\sqrt{2}}\left(
\bigotimes_{i=1}^{N}\left\vert -\right\rangle_{z}^{(i)
}+e^{i\pi(N+1)/2}\bigotimes_{i=1}^{N}\left\vert
+\right\rangle_{z}^{(i)}\right) \; ,\label{ghz1}
\end{equation}
if $\theta(n)=(1+4m)\pi/8$, where $m$ is an arbitrary integer. A
comparison with Eq.~(\ref{theta}) shows that the integers $n$ and
$m$ are related by
\begin{equation}
n=m\frac{\omega^{2}}{4g^{2}}+\frac{\omega^{2}}{16g^{2}}\; ,
\label{nm}
\end{equation}
which is possible only if the (experimentally controllable)
parameter $g^2/\omega^2$ is chosen to be
\begin{equation}
\frac{g^2}{\omega^2}=\frac{1+4m}{16n}\; . \label{cond}
\end{equation}

Since it is difficult in practice to realize $g$ comparable to
$\omega$, we assume $m=0$. Hence Eq.~(\ref{cond}) determines the
value $n_{\rm min}$ (typically larger than $1$) which corresponds to
the minimum preparation time of the GHZ state
\begin{equation}
T_{\mathrm{min}}=\frac{2\pi n_{\rm min}}{\omega}=\frac{\pi\omega}{8g^{2}}.
\label{t}
\end{equation}
The optimal case $n_{\rm min}=1$ could be realized if it were
possible to achieve $g=\omega/4$. The same GHZ state is periodically
generated at later times, with preparation time
$T_{\text{p}}=T_{\mathrm{min}}(1+4m)$.

For both types of qubits, $g$ is proportional to $\sqrt{\omega}$
(since $g$ is proportional to $\eta$, see Eqs.~(\ref{eta}) and
(\ref{ch})). If we assume $g=\xi \sqrt{\omega}$, we obtain
$T_{\mathrm{min}}=\pi/8\xi^{2}$.  Therefore, the preparation time
does not depend on $\omega$. Furthermore, the preparation time
Eq.~(\ref{t}) does not increase with the number of qubits.

If the qubits evolve under the time evolution described by
Eq.~(\ref{u2}) with $\theta(n)=(3+4m)\pi/8$, another $N$-qubit GHZ
state is realized,
\begin{equation}
\left\vert \Psi (T_n) \right\rangle =\frac{1}{\sqrt{2}}\left(
\bigotimes_{i=1}^{N}\left\vert -\right\rangle_{z}^{(i)
}+e^{-i\pi(N+1)/2}\bigotimes_{i=1}^{N}\left\vert
+\right\rangle_{z}^{(i)}\right) \;.\label{ghz2}
\end{equation}
In the following discussion, we focus on the GHZ state
Eq.~(\ref{ghz1}) since it can be prepared in a shorter time.

The treatment discussed up to now is valid if the qubit number $N$
is even. For odd $N$, the single-qubit rotation $U'=\exp(-\pi
J_{x}/2)$ is needed in addition to the time evolution
Eq.~(\ref{u2}). The GHZ state that can be realized for odd $N$ has
the form
\begin{equation}
\left\vert \Psi (T_n) \right\rangle =\frac{1}{\sqrt{2}}\left(
\bigotimes_{i=1}^{N}\left\vert -\right\rangle_{z}^{(i)}+e^{i\pi N
/2}\bigotimes_{i=1}^{N}\left\vert +\right\rangle_{z}^{(i)}\right) \; .
\label{ghz3}
\end{equation}
To conclude: one can prepare an $N$-qubit GHZ state by turning on
the qubit-resonator interaction for a specified time.

For this GHZ state to be useful for quantum information processing,
the preparation time has to be shorter than the quantum coherence time
of the whole system.  In general, a short preparation time results
from a strong qubit-qubit coupling. However, this conflicts with the
weak-coupling condition assumed in many schemes in order to utilize
virtual photon excitation or the rotating-wave approximation. Our
preparation scheme for the GHZ state is based on {\it real
  excitations} of the quantum bus. No weak-coupling condition is
required here. In principle, it can be applied to the `ultra-strong'
coupling regime that the coupling strength between the quantum bus
(i.e. the TLR) and the qubits is comparable to the free system energy
spacing.  Hence it is possible to implement GHZ state preparation in a
very short time. To get an idea of the time scale under realistic
experimental conditions, we now estimate the preparation time using
typical experimental parameters.

Assuming the mutual inductance between qubit and resonator
$M^{(i)}=20$~pH, the self inductance $L=100$~pH, and the resonator
frequency $\omega =1$~GHz leads to $\eta \sim 1.76\times10^{-3}$ for
both types of qubits. For charge qubits, we assume
$E_{J}^{(i)}=14$~GHz, $\Omega^{(i)}=10$~GHz, and that the bias
during the coupling period satisfies
$\sin(\pi\Phi_{e}^{(i)}/\Phi_{0})=0.8$. This leads to a coupling
strength of $g=19.71$~MHz. For flux qubits, we assume a qubit
frequency $\Omega^{(i)}=10$~GHz, $E_{J}^{(i)}=345$~GHz,
$\alpha_0^{(i)}=0.42$, the bias satisfies
$\sin(\pi\Phi_{e}^{(i)}/\Phi_{0})=0.71$, and at this bias,
$d\Delta/d\alpha=112$~GHz. Both $2\alpha_0^{(i)}$ and
$2\alpha_0^{(i)}\cos(\pi\Phi_{e}^{(i)}/\Phi_{0})$ should be within
the interval $(0.6,0.85)$ so that the circuits can always work as
flux qubits both with and without bias. This leads to $g\approx
144$~MHz~\cite{Wang2009}. The coupling strength is much stronger for
flux qubits than charge qubits because of the direct magnetic
coupling to the phase degree of freedom.

Therefore the interaction time to realize a GHZ state is
$T_{\mathrm{min}}=1$~$\mu$s for charge qubits and
$T_{\mathrm{min}}=19$~ns for flux qubits. The preparation time for
flux qubits is much shorter than the coherence time of the TLR which
can be several hundred microseconds. The typical single-qubit
coherence time at the degeneracy point is several microseconds. Hence
in principle, the scheme is able to prepare GHZ states for several
tens of qubits. If the coupling strength can be further increased to
the `ultra-strong' regime in experiment, the preparation of a
multi-qubit GHZ state can be comparable to the time of a single qubit
operation.

\section{Preparation errors}

From the above calculation, it is clear that the essential point to
prepare the GHZ state is to control the length of the dc pulse to
manipulate the flux bias $\Phi_{e}^{(i)}$. In the beginning, the
external magnetic flux $\Phi_{e}^{(i)}$ is set to $n\pi$, all the
qubits and the transmission line resonator are relaxed to their
respective ground states.  Then the interaction between the qubits and
the resonator is turned on by biasing $\Phi_{e}^{(i)}$ away from
$n\pi$ to some appropriate value for a time
$T_{\mathrm{min}}$. Finally, the interaction is switched off by setting
$\Phi_{e}^{(i)}=n\pi$ again, and the multi-qubit GHZ state is
realized. Note that all the qubit biases are modified during the
preparation by the same pulse, therefore all the qubits can share one
control line for the magnetic flux. To accomplish this operation, two
practical issues have to be considered.

\begin{figure}[tp]
\centering \subfigure {\label{fig:istate:a}
\includegraphics[bb=60 435 542 602,
scale=0.5,clip]{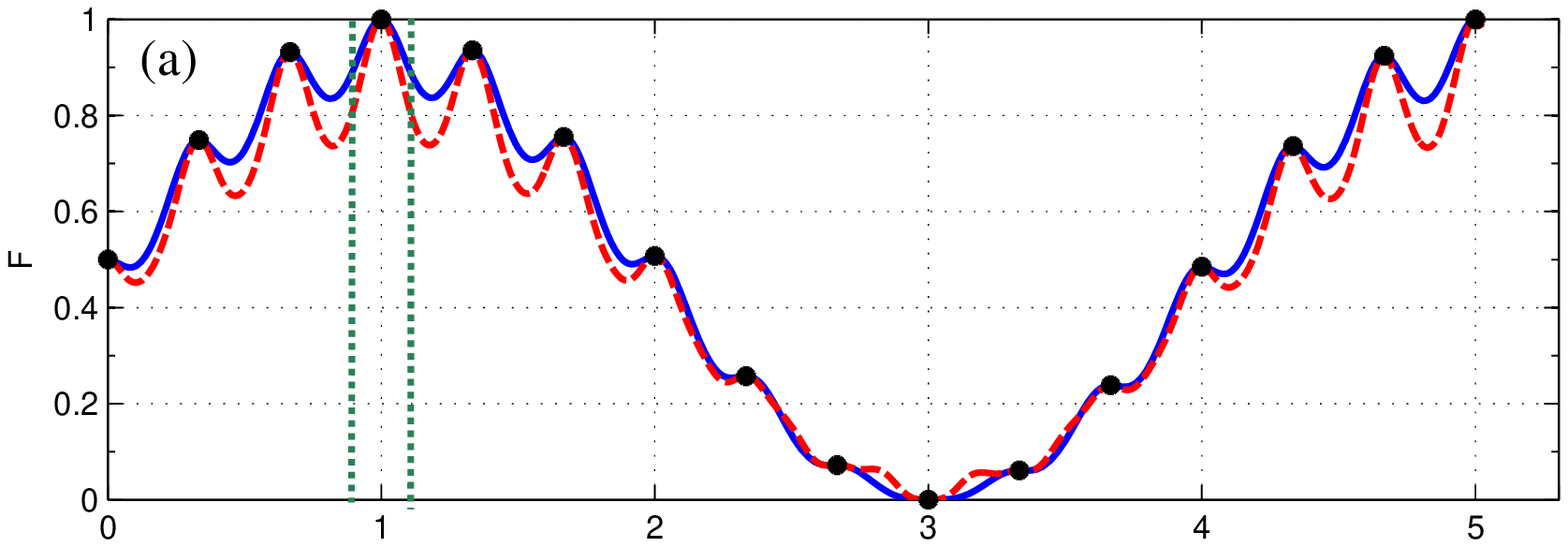}}\\
\subfigure {\label{fig:istate:b}
\includegraphics[bb=60 430 542 602,scale=0.5,clip]{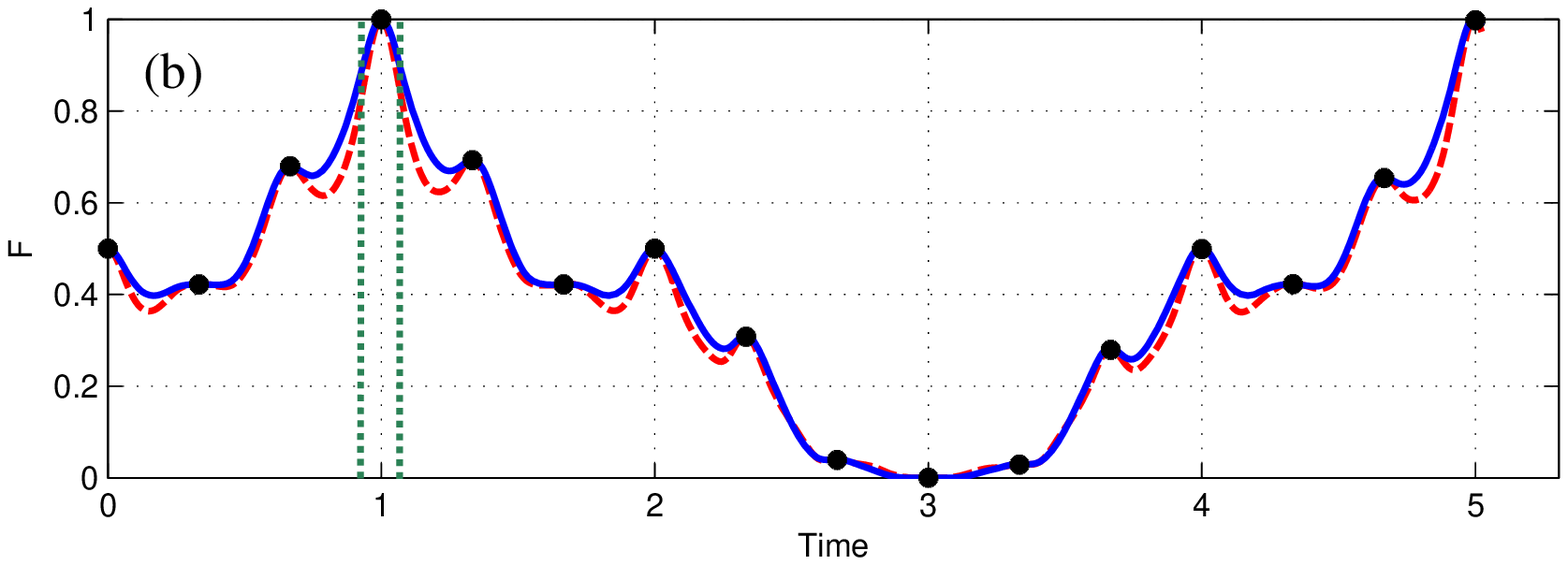}}
\caption{(Color online) Time dependence of the fidelity of the
prepared GHZ state for two different initial resonator states: the
ground state (blue line) and the thermal state (red line) in the
case of (a) two qubits, (b) four qubits. The black dots indicate the
time when the resonator and qubits are effectively decoupled. The
green lines limit the regime in which the fidelity is larger than
$90\%$. The following parameter values were used: qubit frequency
$\Omega^{(i)}=10$ GHz, resonator frequency $\omega =1$ GHz, coupling
strength $g=144$~MHz. The time is given in units of
$T_{\mathrm{min}}$.} \label{fig:istate}
\end{figure}

The first one is the precision of the control of the pulse length to
keep the error acceptable. If the pulse length is not exactly
$T_{\mathrm{min}}$, the state realized is not a GHZ state and this
error can be evaluated by calculating the
fidelity~\cite{Uhlmann1976,Jozsa1994} $F(t)
=\mathrm{Tr}\left[\rho_{\text{GHZ}}\rho_{q}(t) \right]$, where
$\rho_{q}(t)$ is the reduced density matrix of the qubits and
$\rho_{\text{GHZ}}$ is the density matrix of the $N$-qubit GHZ
state. In Fig.~\ref{fig:istate}, the blue curves show the fidelity
of state preparation, the regime with fidelity larger than $90\%$ is
marked by two green dotted lines.  To realize a preparation with
above $90\%$ fidelity, the time control of the pulse should be
precise to around $2.5$~ns in the four qubits case, which is
possible in experiment.

The second problem is the influence of the non-ideal pulse shape. In
the above calculation, we have assumed that a perfect square pulse
can be applied so that $g^{(i)}$ is a constant during the
preparation. However, in experiment the dc pulse generated always
has a finite rise and fall time.  Since the coupling strength
$g^{(i)}$ depends on the bias flux $\Phi_{e}^{(i)}$, the modulation
of the magnetic flux results in a time-dependent coupling strength
$g^{(i)}=g^{(i)}(t)$. If $g^{(i)}$ varies slowly with time (compared
with $e^{-i\omega t}$), the above discussions still hold except that
the decoupling time $T$ at which the qubit-resonator coupling can be
canceled is shifted to satisfy
\begin{equation}
e^{-i\omega T}g^{(i)}(T) -g^{(i)}(0)=0\; .
\end{equation}
A GHZ state is prepared if
\begin{align}
\frac{\pi\omega}{8}=&\int_{0}^{T}dt'\{e^{i\omega t'}
[g^{(i)}(t')g^{(j)}(0)+g^{(i)}(0) g^{(j)}(t')]\nonumber\\
&-2g^{(i)}(t') g^{(j)}(t')\}
\end{align}
for all $i$, $j$.
This means a GHZ state can be realized by dc pulses of finite
bandwidth without introducing additional errors.

Another systematic error appears because the parameter
$g^2/\omega^2$ cannot be controlled with arbitrary accuracy, i.e.,
Eq.~(\ref{cond}) will be satisfied only approximately.
In experiment, $\Phi_e^{(i)}$ is tuned to get the desired value
of $g$; whereas $\omega$ is fixed by the geometry of the device.
Suppose the experimental inaccuracy leads to a
modified value for the coupling strength, $g(1+\delta)$, where
$\delta$ quantifies the magnitude of error.
Hence the prepared state deviates from the GHZ state. The fidelity of the
prepared state depends on $\delta$ as
\begin{equation}
F(\delta)=\left\vert \left\langle \Psi(T)
|\text{GHZ}\right\rangle\right\vert ^{2}= \frac{1}{2^{2N}}\left\vert
\sum_{r=0}^{N}C_{N}^{r}e^{i\frac{\pi}{2} (\delta^2+2\delta) \left(
\frac{N}{2}-r\right) ^{2}}\right\vert ^{2}\; .
\label{f_delta}
\end{equation}
where $C_{N}^{r}=N!/(r!(N-r)!)$ is the binomial coefficient. This
expression is valid for even $N$. For odd $N$,
the fidelity turns out to be given by Eq.~(\ref{f_delta}) with $N \to N+1$.
\begin{figure}[bp]
\begin{center}
\includegraphics[bb=101 320 486 515, scale=0.6, clip]{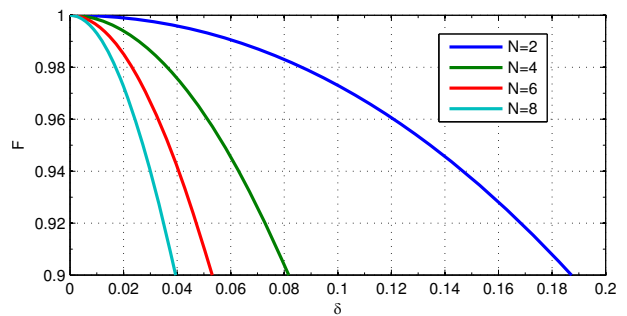}
\end{center}
\caption{(Color online) Dependence of the fidelity $F$ on the error
of the coupling coefficient $\delta$.
The curves correspond to $N=2$ (top), 4, 6, and 8 (bottom).}
\label{fig_lambda}
\end{figure}
Figure~\ref{fig_lambda} shows that the fidelity decreases as the error
in the coupling coefficient increases. In the case of a 4-qubit GHZ
state, a fidelity of $98\%$ can be achieved if the error in $g$ is
within $3\%$. However, as the number of qubits increases, the fidelity
drops more rapidly.  Hence a more precise control of the flux bias is
required to realize many-qubit GHZ states.

\section{Error caused by decoherence}
An important advantage of our proposal is that the state preparation
is independent of the initial state of the resonator. In general, it
is not easy to prepare the system to be exactly in the ground state.
For example, at typical dilution fridge temperatures, say $50$~mK,
there is a non-negligible probability ($30\%$) for the first excited
state of a $1$~GHz resonator to be occupied. This problem is less
severe for the qubits since their energy scale is much higher.
Therefore a scheme which is insensitive to the initial state is
desirable.

Figure~\ref{fig:istate} shows the fidelity of the prepared GHZ state
for two different initial states of the resonator (the ground state
and the thermal state at $50$~mK). Although the time evolutions are
different in general, the fidelities at the decoupling time $T_{n}$
(indicated by black dots in the figures) are the same. This can be
explained from Eq.~(\ref{u1}), at times $T_{n}$, only the first term
of Eq.~(\ref{u1}) is kept, i.e., the qubits and resonator are
decoupled. No matter what the initial state of the resonator is, at
these times, the resonator has evolved back to its initial state.
This means the GHZ state preparation is not influenced by the
initial state, or in other words, the preparation is insensitive to
the decoherence that occurred before the interaction was switched
on.

\begin{figure}[tbp]
\centering
\subfigure {\label{fig:deco:a}
\includegraphics[bb=52 436 541 602,scale=0.5,clip]{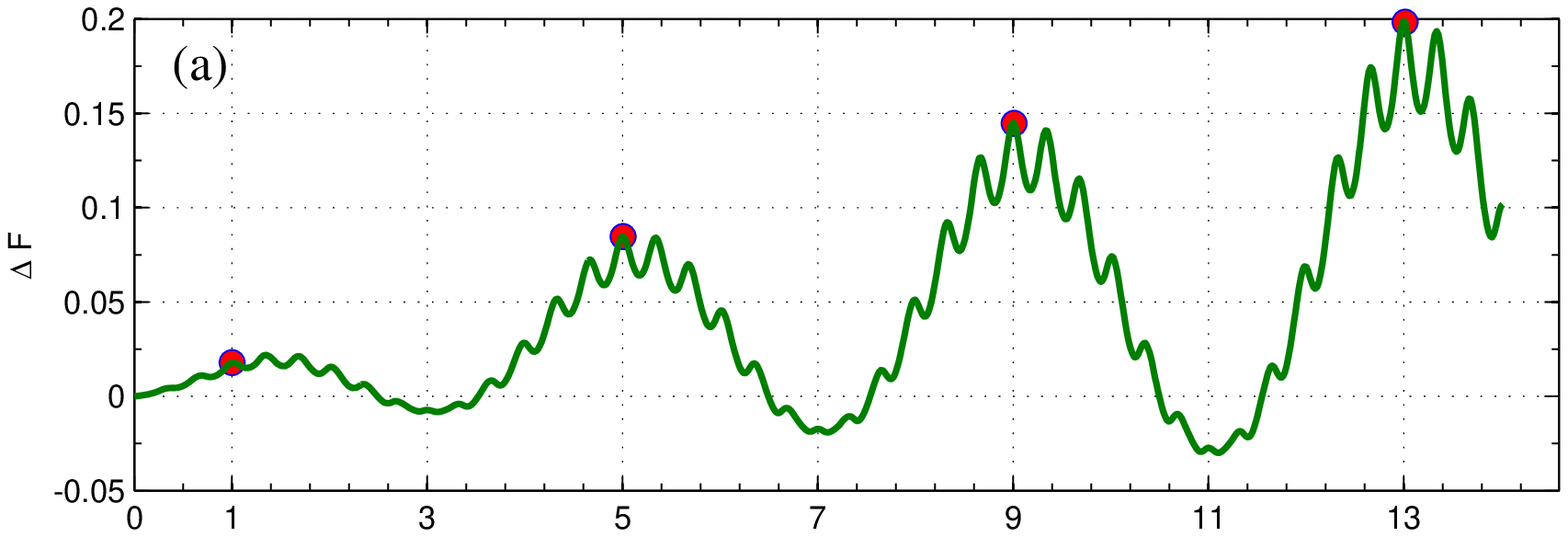}}\\
\subfigure {\label{fig:deco:b}
\includegraphics[bb=52 429 541 602,scale=0.5,clip]{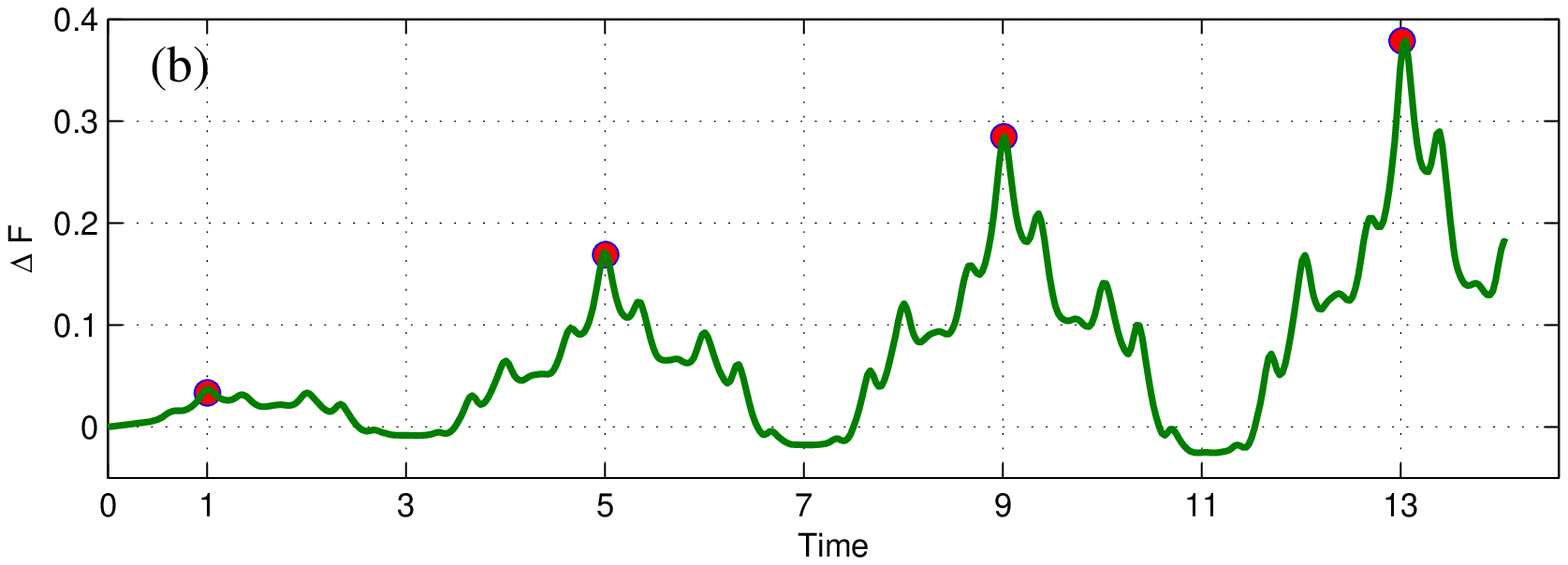}}
\caption{(Color online) Time dependence of the error due to
decoherence. $\Delta F$ is the difference of the fidelity of the
prepared GHZ state with/without environment decoherence for (a) 2
qubits (b) 4 qubits. The red dots mark the times at which the GHZ state is
prepared. The following parameter values were used: qubit frequency
$\Omega^{(i)}=10$ GHz, resonator frequency $\omega =1$ GHz, coupling
strength $g=144$~MHz, qubit decay rate $\gamma =10$ MHz, and
resonator decay rate $\kappa =0.5$ MHz. The time is given in units of
$T_{\mathrm{min}}$.} \label{fig:deco}
\end{figure}

But the decoherence during the operation certainly changes the final output
state. In general, environmental fluctuations induce both dephasing and
relaxation to the system. Since the qubits are all biased at the degeneracy
point, the strong dephasing effect due to 1/f noise is largely suppressed.
Thus we can use a master equation which only includes relaxation as damping
instead of the unitary operator Eq.~(\ref{u1}) to fully characterize the
time evolution
\begin{equation}
\dot{\rho}(t)=-i[H,\rho(t)]+ \mathcal{L}_{Q}\rho(t)
+\mathcal{L}_{R}\rho(t)\; , \label{m1}
\end{equation}
where $\rho(t)$ is the density matrix of the system (qubits + resonator)
in the interaction picture and $\mathcal{L}_{R}$ represents the
decoherence of the resonator
\begin{align}
\mathcal{L}_{R}\rho =&\frac{\kappa}{2}(N_{\text{th}}+1) (2a\rho
a^{\dag}-a^{\dag}a\rho -\rho a^{\dag}a)\nonumber\\
&+\frac{\kappa}{2}
N_{\text{th}}(2a^{\dag}\rho a-aa^{\dag}\rho -\rho aa^{\dag}) \; .
\end{align}
Here, $\kappa$ is the resonator decay rate and
$N_{\text{th}}=(\exp(\omega /k_{\text{B}}T) -1)^{-1}$ the average
number of photons in the resonator. Finally, $\mathcal{L}_{Q}$
represents the decoherence of the qubits
\begin{equation}
\mathcal{L}_{Q}\rho =\frac{\gamma}{2}(2\tilde{\sigma}_{-}\rho
\tilde{\sigma}_{+}-\rho
\tilde{\sigma}_{+}\tilde{\sigma}_{-}-\tilde{\sigma}_{+}\tilde{\sigma}_{-}\rho)\;
,
\end{equation}
where $\gamma$ is the qubit decay rate and $\tilde{\sigma}_\pm$ are
written in the diagonal basis of $\sigma_x$. The quality factor of a
TLR can be as high as than $10^{6}$. The qubit $T_1$-time at the
degeneracy point is several $\mu$s at most in present experiment. To
be on the safe side, we assume for the resonator $Q=2\times 10^{3}$,
$\kappa =0.5$~MHz, and for the qubit $T_{1}=100$~ns, i.e., the decay
rate is $\gamma =10$~MHz. Here we neglect excitations of the qubit
since its energy spacing is much larger than the thermal
fluctuation. To investigate the influence of decoherence, we compare
the fidelity to prepare a GHZ state with/without decoherence. The
result is shown in Fig.~\ref{fig:deco} where the difference of the
two fidelities $\Delta F=F-F_{\text{d}}$ (where $F_{\text{d}}$ is
the fidelity in the presence of decoherence) is plotted as a
function of time. The red dots mark the difference at the GHZ
preparation times $T_{p}$. Obviously, the error due to decoherence
increases with time. As we analyzed in the previous section, the
preparation time is much shorter than the decoherence time.
Therefore the error is still quite small at the minimum preparation
time $T_{\mathrm{min}}$ (indicated by the first dot): the error
caused by decoherence is around $3.7\%$ in the 4-qubit case.

\section{Discussion and Conclusion}
In the above discussion, for simplicity, we assumed that the qubits
only interact with a single mode of the resonator. However, since we
did not invoke the rotating wave approximation in our calculation,
higher modes of the TLR~\cite{Blais2004,Goppl2008} will also
contribute to the coupling. Therefore, the interaction
Eq.~(\ref{hi}) should include a sum over multiple modes whose
frequencies are below a cut-off $\omega_{\text{c}}$. The cut-off is
determined by a number of practical issues, e.g., by the
superconducting gap, or the fact that the resonator is not strictly
one-dimensional~\cite{Blais2004}.  The time evolution including
higher modes is of the same form as Eq.~(\ref{u1}) but
includes a product over all the relevant modes. Neglecting the small
nonlinear effect due to output coupling, the frequencies of all
higher modes are multiples of the frequency of the fundamental mode,
$\omega_{\tilde{n}}=\tilde{n}\omega$ and
$\omega_{\text{c}}=\tilde{n}_{\text{c}}\omega$, all the coupling
coefficients between the qubits and different modes of the TLR,
$B_{i,\tilde{n}}(t)=
ig^{(i)}_{\tilde{n}}(e^{-i\omega_{\tilde{n}}t}-1)/\omega_{\tilde{n}}$
are still zero for $t={T}_{n}=2n\pi/\omega$. Here
$g^{(i)}_{\tilde{n}}\equiv g^{(i)}(\omega_{\tilde{n}})$. Hence the
only correction to our scheme is including a sum over all the
relevant modes in the definition of $A_{ij}(t)$ in Eq.~(\ref{ca}),
\begin{equation}
A_{ij}(t)=\sum_{\tilde{n}=1}^{\tilde{n}_c}
\frac{g^{(i)}_{\tilde{n}}g^{(j)}_{\tilde{n}}}{\omega_{\tilde{n}}}t\;.
\end{equation}
For low excitation modes whose wave lengths are still much larger
than the qubit dimension, the homogeneous coupling assumption still
approximately valid, i.e., $g^{(i)}_{\tilde{n}}\equiv
g_{\tilde{n}}$. For example, considering $\tilde{n}_c=10$ for a $10$
cm transmission line, around the center there is a $0.32$~mm-long
region where the magnetic field varies within $5$~\%. The distance
between the center of two qubits is roughly $10\mu $m. This means up
to around 30 qubits are coupled to the resonator approximately homogeneously.
One can also tune $\Phi_{e}^{i}$ to further compensate the slight
inhomogeneity. The correction to the time evolution Eq.~(\ref{u2})
can be simply written as $\theta(n) =(2\pi
n)g^{2}(\tilde{n}_{\text{c}}/2)/\omega^{2}$. Therefore, the effect
of the higher excitation modes actually amounts to increasing the
coupling coefficient $g\rightarrow g\sqrt{\tilde{n}_{c}/2}$, which
helps to reduce the operation time.

The electric field of the higher modes has little effect on the flux
qubits but will change the voltage bias of the charge qubits and
couple to their $\sigma_z$ component. However, at the degeneracy
point, where the free Hamiltonian is proportional to $\sigma_x$, these
coupling terms are rapidly oscillating and are expected to have a
small effect on the system. However, the situation is less
advantageous than for flux qubits, and higher modes should be
suppressed by choosing high fundamental mode frequencies in the
charge-qubit case.

For a small number of qubits, an (lumped) LC circuit can also be
used as a quantum bus~\cite{Chiorescu2003,Johansson2006} to generate
a GHZ state by following our scheme. In this case, only one single
mode contributes.

In conclusion, we have proposed a scheme to prepare an $N$-qubit GHZ
state in a system of superconducting qubits coupled by a transmission
line resonator. We have analyzed the preparation scheme for both
charge qubits and flux qubits.  With this method, a multi-qubit GHZ
state can be prepared within the quantum coherence time. In the case
of flux qubits that is especially favorable, the preparation time is
two orders of magnitude shorter than the qubit coherence time. The
preparation time can be reduced further if the coupling strength is
increased to the ultra-strong coupling regime, where the coupling
strength is comparable to the free qubit Hamiltonian. The preparation
scheme is insensitive to the initial state of the resonator and robust
to operation errors and decoherence. The coupling can be switched by
dc pulses of finite rise and fall times without introducing additional
errors. In addition, the scheme described in this paper utilizes a
linear coupling which is intrinsically error-free if proper dc control
is achieved. Due to all these advantages, this proposal could be a
promising candidate for GHZ state generation in systems of
superconducting qubits.

\section{Acknowledgement}

The authors acknowledge helpful discussions with A. Wallraff and Yong
Li. This work was partially supported by the EC IST-FET project
EuroSQIP, the Swiss SNF, and the NCCR Nanoscience.

%\bibliography{ghzgen}

\end{document}